\documentclass[
reprint,
superscriptaddress,
amsmath,amssymb,
aps,
prb,
longbibliography
]{revtex4-1}

\usepackage{graphicx,epsf}
\usepackage{float}
\usepackage{xcolor}
\usepackage{bm}

\begin{document}

\title{
Geometry effects in topologically confined bilayer graphene loops
}

\author{Nassima Benchtaber}
\affiliation{Institute of Interdisciplinary Physics and Complex Systems IFISC 
(CSIC-UIB), E-07122 Palma, Spain} 
\author{David S\'anchez}
\affiliation{Institute of Interdisciplinary Physics and Complex Systems IFISC 
(CSIC-UIB), E-07122 Palma, Spain} 
\affiliation{Department of Physics, University of the Balearic Islands, 
E-07122 Palma, Spain}
\author{Lloren\c{c} Serra}
\affiliation{Institute of Interdisciplinary Physics and Complex Systems IFISC 
(CSIC-UIB), E-07122 Palma, Spain} 
\affiliation{Department of Physics, University of the Balearic Islands, 
E-07122 Palma, Spain}

\begin{abstract}
We investigate the electronic confinement in bilayer graphene 
by topological loops of different shapes. These loops are created 
by lateral gates acting via gap inversion on the two graphene sheets.
For large-area loops
the spectrum is well described by a quantization rule depending only on the loop perimeter. For small sizes, the spectrum depends on the loop shape. 
We find that zero-energy states exhibit a characteristic pattern that strongly depends on the spatial symmetry. We show this by considering loops of higher to lower symmetry
(circle, square, rectangle and irregular polygon). 
Interestingly, magnetic field causes valley splittings of the states, an asymmetry between energy reversal states, flux periodicities and
the emergence of persistent currents.
\end{abstract}

\maketitle

\section{Introduction}

One of the most fundamental predictions of Quantum Mechanics is the existence of confined or bound states which are eigenstates of the Hamiltonian describing a particular system~\cite{Landau}. It is well known that the Schr\"odinger
equation with a spatially dependent potential yields 
in many cases bound eigenstates located around the absolute potential minima. This potential binding 
mechanism around minima is crucial in many natural physical systems like, e.g., electrons in atoms and molecules~\cite{Bransden};
and it is also behind the  formation of artificial bound states by potential gating with microelectrodes,
as in semiconductor quantum dots~\cite{Ihn}. In bilayer graphene (BLG) with 
Bernal stacking, the
material we consider in this
work, a similar potential confinement mechanism exists but it is related to the potential 
difference $V_a$ applied to 
split dual
gates on the two graphene sheets; electrons are bound to the regions where $V_a$ is lowest. These regions of lowest $V_a$ can take different shapes, such as circular dots and rings for instance, and they have been intensively studied both theoretically and
experimentally~\cite{Trau07,Pereira07,Recher09,Recher10,Gon10,Gonb11,Ore13,Dacosta14,Eich18,Kurzmann19,Cle19,Banszerus20,Banszerus21}.
The technological interest of BLG nanostructures is twofold. First, they could serve as scalable spin qubits with long decoherence
times due to the weak spin-orbit interaction in graphene~\citep{min06,yao07,gmit09}.
Second, in combination with magnetic fields they could be utilized for valleytronic
operations thanks to the manipulation of the valley degree of freedom~\cite{Men15,Over18,Kraf18}.

The possibility of an even more robust type of confinement  in BLG was suggested in Ref.~\onlinecite{Mar08}.
This is a qualitatively different confinement, of a topological character, emerging  near the domain wall that separates two regions where
$V_a$ %, the potential difference between the two grahene layers,
changes sign.
The confined states are intrinsically 1D-like, with a characteristic transverse decay length, and they propagate along the domain wall with locked relative orientations of 
momentum and valley pseudospin. In the absence of valley mixing potentials states for the K and K' valleys propagate along the wall, without backscattering, showing opposite chiralities.
This property is similar to the helical states within the quantum spin Hall effect~\cite{kan05,ber06,kon07}
if we replace spin with the valley pseudospin of graphene.

\begin{figure}[b]
\begin{center}
\includegraphics[width=0.5\textwidth, trim = 1.5cm 0.5cm 4cm 0.5cm, clip]{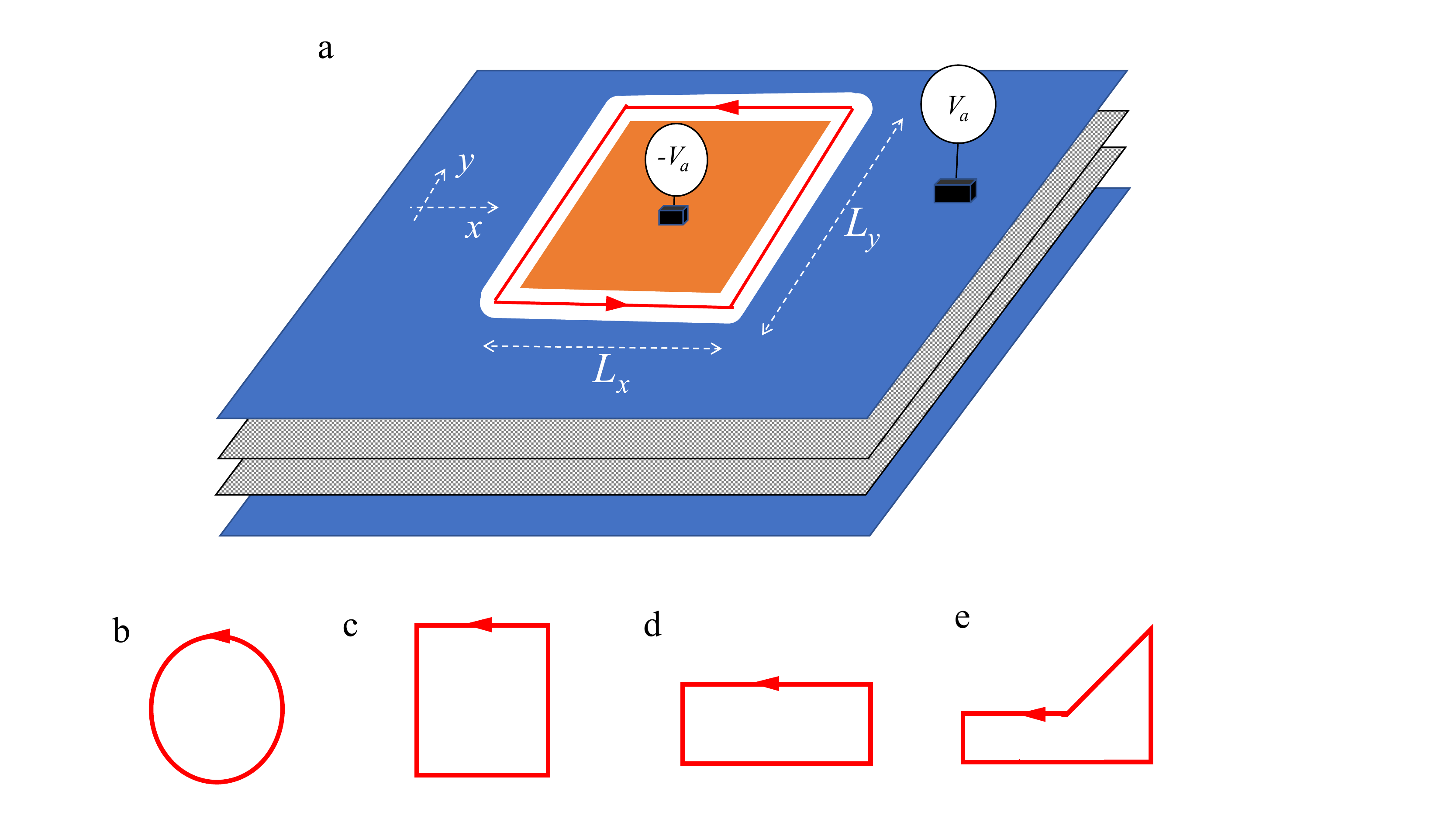}
\end{center}
 \caption{a) Scheme of a topological loop showing  the bilayer graphene sheets (gray) and lateral gates with the applied gate potentials $\pm V_a$ (orange and blue, respectively). Two identical lower gates, hidden behind the lower graphene sheet, have the opposite potentials of the 
 corresponding top gates. The white 1D region between orange and blue gates hosts the topological loop state on the graphene sheets, with counterpropagation for the two valleys. A red arrow is indicating the circulation for only one of the two valleys.
 b)-e) Loop shapes considered in this work,
 from highest to lowest symmetry:
 circle, square, rectangle and irregular polygon.
 }
\label{Fig1}
\end{figure}

Now, let us suppose that the external gates are arranged in such a way that the domain wall closes into itself.
A schematic representation is shown in Fig.\ \ref{Fig1}a. Then, a loop forms supporting electronic bound states~\cite{xavier10} and
the following question naturally arises. What is the effect of different loop shapes onto the energy spectrum of these bound states?
Being 1D-like, topologically confined sates  in BLG are expected to essentially depend on the loop perimeter
unlike the trivial potential confinement, which depends both on the surface and the shape of the confining region.
Nevertheless, it is important to notice that the domain-wall states 
depicted with a red line in Fig.\ \ref{Fig1}a are in reality quasi-1D with a finite spatial width. Therefore,
the discrete energy spectrum will be in general a function of the system's geometrical symmetries, the region
aspect ratio, etc. Below, we present a detailed investigation for topological loops of different shapes and symmetries.
First, we show with the aid of an exactly solvable model that the energy spectrum is uniquely determined
by the loop perimeter provided that the system size is large. Yet, for smaller loops finite size effects become
important and the spectrum depends on the particular geometrical structure. This dependence is better seen in the
sequence of zero energy crossing and anticrossings as the perimeter increases.

%In this work we study confined states in BLG topological loops of different shapes, focussing on the evolution of the 
%spectrum of eigenstates as the perimeter is increased. A simplified model based on a quantization rule along the perimeter, similar to Bohr's atomic orbit model, is shown to describe well large size loops. In smaller loops we discuss characteristic differences for the spectra of different shapes as a function of perimeter which are induced 
%by quantum finite size (length) effects.
%The most remarkable being the presence of a regular sequence of zero energy 
%crossings in rings, a sequence of alternated crossings and anticrossings in squares, and only anticrossings in rectangles and more irregular structures. In all cases, as the perimeter %increases, if present the anticrossings tend to smoothly vanish, converging to the mentioned analytic model behavior.

In presence of a magnetic field $B$, the dispersion is affected by the flux piercing the topological loop. This
creates an energy splitting for the two valleys and an asymmetry with respect to energy inversion ($E\to -E$) 
in the spectra as a function of perimeter. The $B$-dependence
of the spectra with a fixed perimeter and shape is characterized by magnetic periodicities of Aharonov-Bohm 
type. Therefore, the loops host persistent currents at finite fields that, similarly to the $B=0$ spectra, 
exhibit geometry dependent features in the small size limit, 
showing that currents would serve as an excellent
tool to probe topologically confined states in BLG nanostructures.
%We stress that BLG topological loops with circular symmetry were discussed in Ref.\ \onlinecite{xavier10} in absence of magnetic field. Here we extend the discussion to include different shapes and magnetic effects.

\section {Theory}

\subsection{A Quantized-perimeter model (QPM)}

An analytic model of topological loops in BLG can be 
devised utilizing the results of Ref.\ \onlinecite{Mar08} for 
 a straight domain wall with an abrupt
transition $+V_a$ to $-V_a$ (a kink). The analytic relations between linear momentum
$p$ along the kink and energy $E$ of the two topological branches, corresponding to the two valleys $(K,K')$, were given in 
Ref.\ \onlinecite{Mar08} as $p=f_{1,2}(E)$, with
\begin{equation}
\label{eq1}
f_{1,2}(E) = 
\frac{
-E\pm \frac{V_a}{\sqrt{2}}
}{
\left(
\mp E + V_a \sqrt{2}\,
\right)^{1/2}
}\,
\frac{\sqrt{t}}{\sqrt{2} v_F}
\; ,
\end{equation}
where the Fermi velocity $\hbar v_F= 660\, {\rm meV}\,{\rm nm}$  and the interlayer coupling $t=380\,{\rm meV}$ are BLG intrinsic parameters.

In a closed loop of perimeter ${\cal P}$ we can assume that the tangent momentum $p$ is quantized,
such that an integer number of wavelengths must fit into ${\cal P}$,
\begin{equation}
p \rightarrow p_n \equiv \hbar\frac{2\pi n}{\cal P}\;,\quad 
n=\pm 1,\pm 2, \dots\; ,
\end{equation}
where negatives $n$'s represent negative momenta 
(i.e., opposite propagation)
and the perimeter quantization in terms of wavelength
$\lambda$ reads ${\cal P}=|n| \lambda$. We can also add a magnetic field, represented by a vector potential $\vec{A}$, and write the circulation integral 
\begin{equation}
\int_{\cal P}{ dl
\left(
p_n + e A_t 
\right)
} 
=
f_{1,2}(E)\int_{\cal P}{dl}\; ,
\end{equation}
where $A_t$ is the  component of the vector potential tangent to the kink.
Noticing that $\int_{\cal P}dl={\cal P}$ and $\int_{\cal P} dl\, A_t=\Phi$, $\Phi$ being the magnetic flux across the loop, we find an implicit condition from which one can derive the bound state energy $E$ for a given loop perimeter, magnetic flux and principal quantum number $n$: %Namely, the condition reads
\begin{equation}
\label{eq4}
 f_{1,2}(E) = \frac{2\pi \hbar}{\cal P} \left(\frac{\Phi}{\Phi_0}+ n \right)\, ,
 \end{equation}
where $\Phi_0=h/e$ is the flux quantum.

Equation (\ref{eq4}) is our quantized-perimeter model (QPM)
for bound states in BLG topological loops. As anticipated, when $\Phi=0$ the relation depends only on the perimeter and is totally independent of the loop geometry. At finite 
fields, however, the bound-state energies depend on the loop surface through the Aharonov-Bohm flux ratio
$\Phi/\Phi_0$. We expect the QPM to be reliable for large
enough loops, when different parts of the loop do not interfere.
In small loops, the traversal extension of the topological states becomes comparable to the distances inside the loop and the QPM breaks down.
Below, we investigate deviations from the QPM using an exact approach for BLG within the continuum limit.

\subsection{A quantum 2D model (Q2DM)}
\label{Q2DM}

%An improved description of the bound states in topological loops can be obtained from
The low-energy Hamiltonian describing the states that are formed in 
two-dimensional (2D)
BLG nanostructures for energies near the Dirac points reads \cite{Mcan13,rozhkov16}
\begin{eqnarray}
 H &=& v_F \left(p_x- \hbar  \frac{y}{l_z^2}\right) \tau_z \sigma_x
 + v_F\, p_y \sigma_y \nonumber\\
 &+& \frac{t}{2}\, \left(\,\lambda_x \sigma_x +\lambda_y\sigma_y\,\right) 
 + V_a(x,y)\, \lambda_z\; ,
\label{eq5}
 \end{eqnarray}
 where $\sigma_{x,y,z}$,
$\tau_{x,y,z}$ and 
$\lambda_{x,y,z}$ are Pauli matrices for the sublattice,
valley and layer pseudo spins, respectively. A topological loop forms from a space dependent function $V_a(x,y)$ that, 
as sketched in Fig.\ \ref{Fig1}a, takes the constant values $+V_a$ ($-V_a$) outside (inside) the loop.
In our numerical simulations,
we consider a smooth spatial transition with a diffusivity\cite{ben21} $s$  
to mimic a realistic experiment~\cite{Li16,Chen20}.
Details of how we model the closed loops of different shapes are given in the Appendix.
Finally, the  field strength 
is included in the magnetic length $l_z=\sqrt{\hbar/eB}$.

We numerically look for the eigenstates  of the Hamiltonian given by Eq.\ (\ref{eq5}) using finite difference discretization of the $xy$ plane in a square grid~\cite{arpack}.
%and the ARPACK matrix eigensolvers.
The spurious solutions due to Fermion doubling~\cite{Susskind77,Nielsen81,Lewe12} have been filtered out as in Ref.\ \onlinecite{ben21}
by coarse graining the wave functions.
Our method can handle any symmetry of the loop, 
in contrast to radial grids which are of smaller dimension and more efficient computationally, but can only describe radially symmetric structures by construction.
In all the cases treated below we have checked that good convergence with the grid size is reached.

The Hamiltonian Eq.\ (\ref{eq5}) fulfills a chiral symmetry given by operator 
${\cal C}=\sigma_x\tau_x\lambda_y$ 
relating eigenstates of opposite energies 
${\cal C}H{\cal C}=-H$ with ${\cal C}^2=1$.
In absence of magnetic field it also 
fulfills time reversal symmetry $\Theta=i\tau_y {\cal K}$, with ${\cal K}$ 
representing complex conjugation,
relating states of the same energy
$\Theta H \Theta = -H$ and with $\Theta^2=-1$.

\section{Results}

\begin{figure}[t]
\begin{center}
\includegraphics[width=0.45\textwidth, trim = 0.5cm 17.5cm 1.cm 2.5cm, clip]{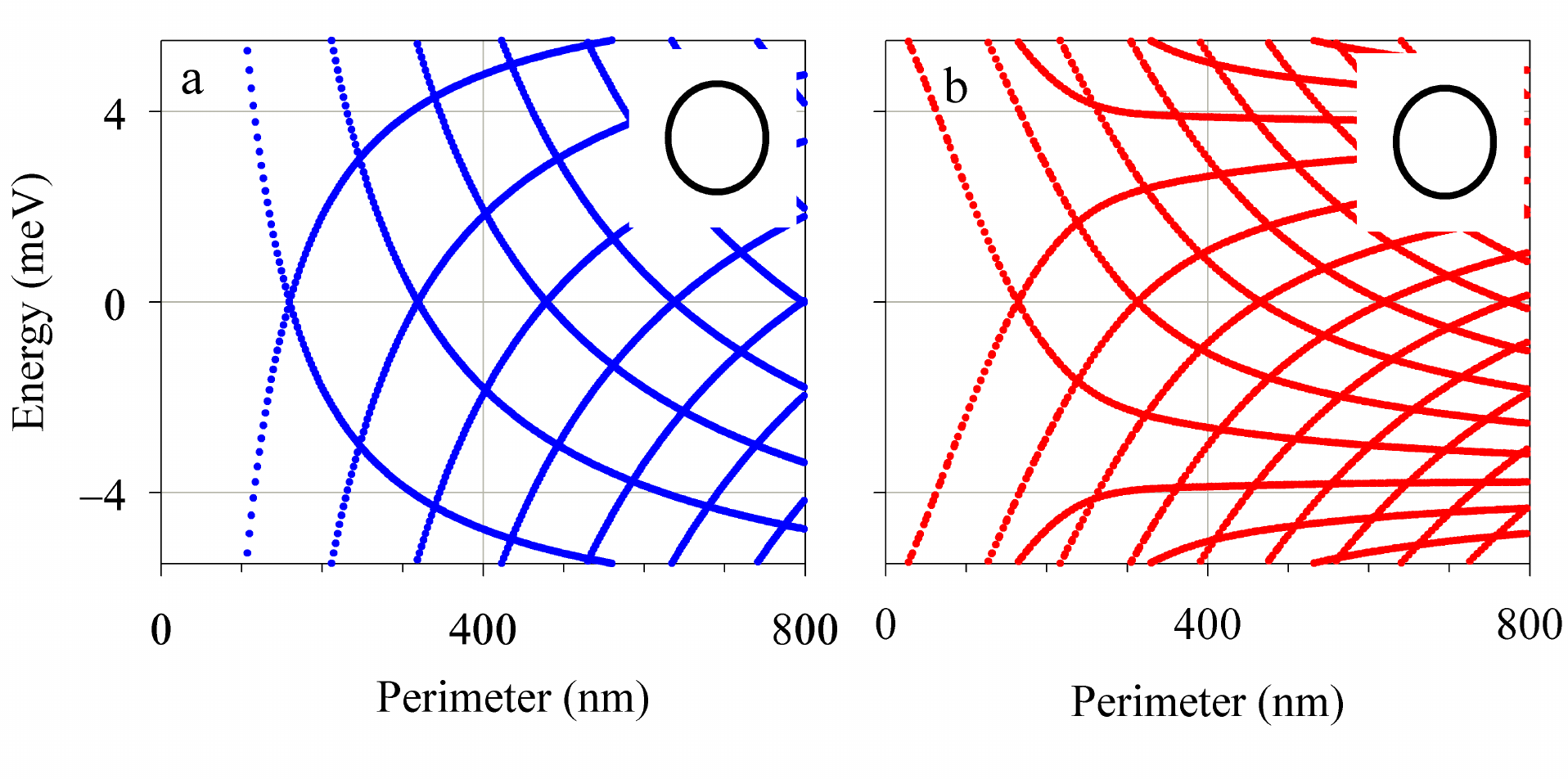}
\end{center}
 \caption{QPM (a) and Q2DM (b) energy spectrum for a circle as a function of its perimeter at zero magnetic field.
 Parameters: wall potential $V_a=10\, {\rm meV}$ and smoothness $s=12.5\, {\rm nm}$.
}
\label{Fig2}
\end{figure}

Figure \ref{Fig2} compares the bound state energies in the analytical (QPM) and numerical (Q2DM) approaches as a function of the loop perimeter
when the structure is a circle. Qualitatively, both approaches nicely agree.
At zero energy,
the QPM predicts a sequence of uniformly spaced branch crossings, with spacing given by the zero-energy wavelength
\begin{equation}
 \lambda_0 = 2^{5/4}\frac{2\pi\,\hbar v_F}{\sqrt{V_a t}}\;, 
\end{equation}
which for $V_a=10\;{\rm meV}$ yields $\lambda_0\simeq 160\; {\rm nm}$. We then find an excellent correspondence with the zero energy crossings obtained numerically in the Q2DM. 
Physically, %the branch crossings are easily understood for the case of the circular ring,
each branch corresponds to an angular momentum and, therefore, can produce crossings with additional branches of different angular momenta. For energies departing from zero, the main difference between panels a and b of Fig.\ \ref{Fig2} is that the energy branches are more densely
spaced in Q2DM than QPM. This difference can be attributed to the small diffusivity $s=12.5\, {\rm nm}$ used in the QD2M. A smoother 1D kink is known to 
reduce the spacing of energy branches~\cite{Zarenia11,ben21}.

\begin{figure}[t]
\begin{center}
\includegraphics[width=0.45\textwidth, trim = 0.5cm 10.5cm 1.6cm 2.5cm,clip]{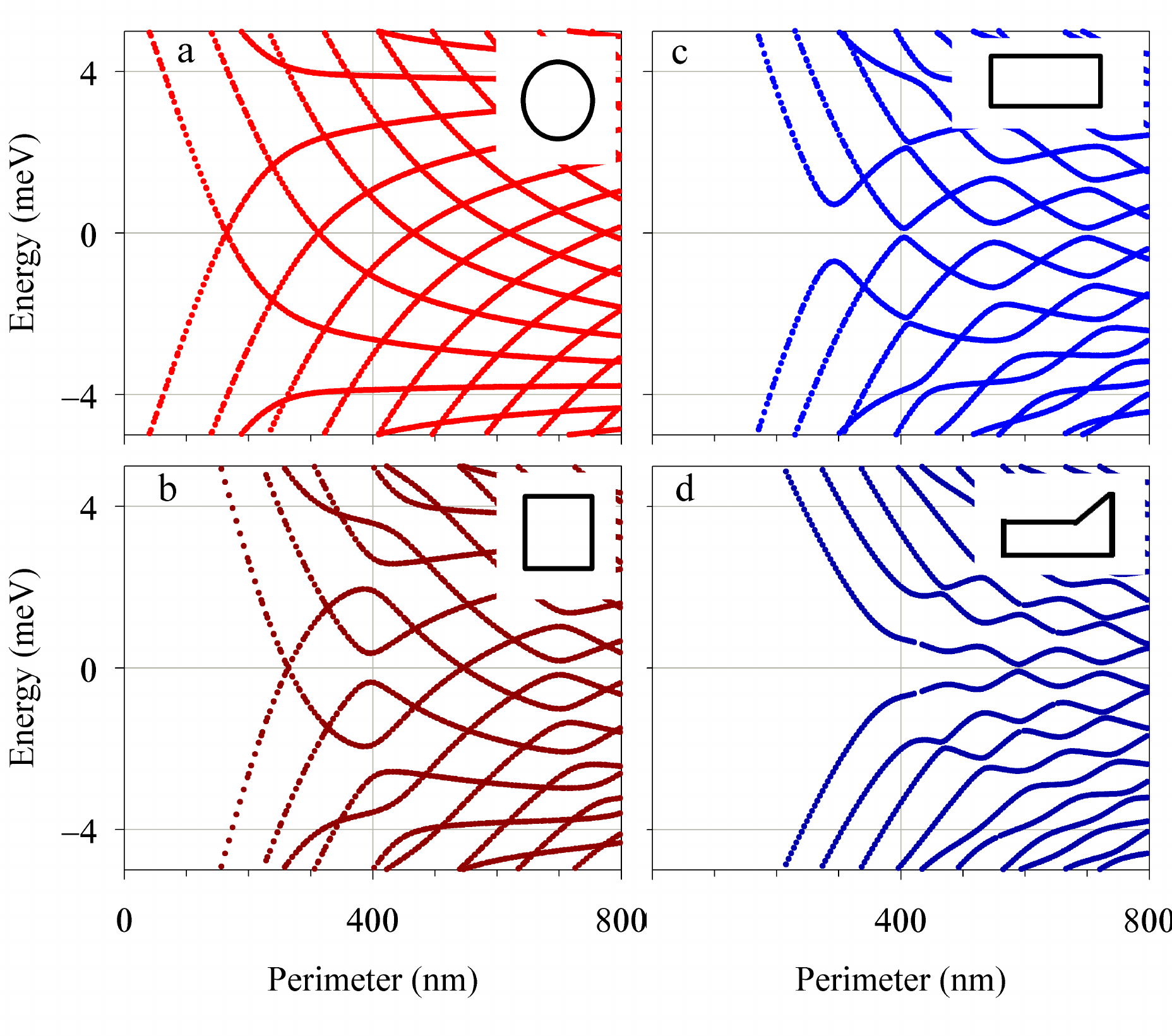}
\end{center}
 \caption{Q2DM energy spectra as a function of perimeter 
 for circle (a), square (b), rectangle (c) and irregular polygon (d).
 Parameters as in Fig.~\ref{Fig2}.
}
\label{Fig3}
\end{figure}

In Fig.\ \ref{Fig3} we show the Q2DM energy spectra for different loop shapes as a function of their perimeters.
Remarkably, the spectrum for noncircular loops show
anticrossings of the branches. The square (b) shows an alternation of zero energy crossings and anticrossings as the perimeter increases. Both the rectangle (c) and the irregular polygon (d) show zero energy anticrossings in all cases, although the size of the anticrossing gap is not uniform. 
There is a general tendency for all shapes to decrease the anticrossing gaps for large perimeter, making the spectra shape independent in this limit.

The fact that, in addition to the circle, only the square 
shows crossings for some branches can be understood as a 
symmetry consequence. Inspired by the QPM, the branch crossing is possible in a square when, simultaneously, two conditions are fufilled
\begin{equation}
\label{eq7}
\left\{
\begin{array}{l}
{\cal P} = n_1 \lambda_0\; ,\\
e^{ik_0L_x}=\pm i\Rightarrow {\cal P}=(2 n_2+1)\lambda_0\; ,
\end{array}
\right.\quad
n_1,n_2 \in \mathbb{N}\; ,
\end{equation}
where $k_0=2\pi/\lambda_0$ is the mode wavenumber. 
Only with $n_1$ odd the crossing is then possible, as 
indeed obtained in Fig.\ \ref{Fig3}b.
The second condition in Eq.~(\ref{eq7}) corresponds to the requirement that a translation along the perimeter by the length of a side changes the state in
a $\pm i$ phase, which occurs when the total number of wavelengths to distribute
in the four square sides is odd $n_1=2n_2+1$.
In a rectangle one should replace $L_x$ by $L_x+L_y$ in this second condition, which then becomes not compatible with the first one and no crossing is therefore allowed.
For the irregular shape (d), the lack of any symmetry 
is responsible for the absence of crossings in the spectrum. 

\begin{figure}[t]
\begin{center}
\includegraphics[width=0.45\textwidth, trim = 0.5cm 17.5cm 1.cm 2.5cm,clip]{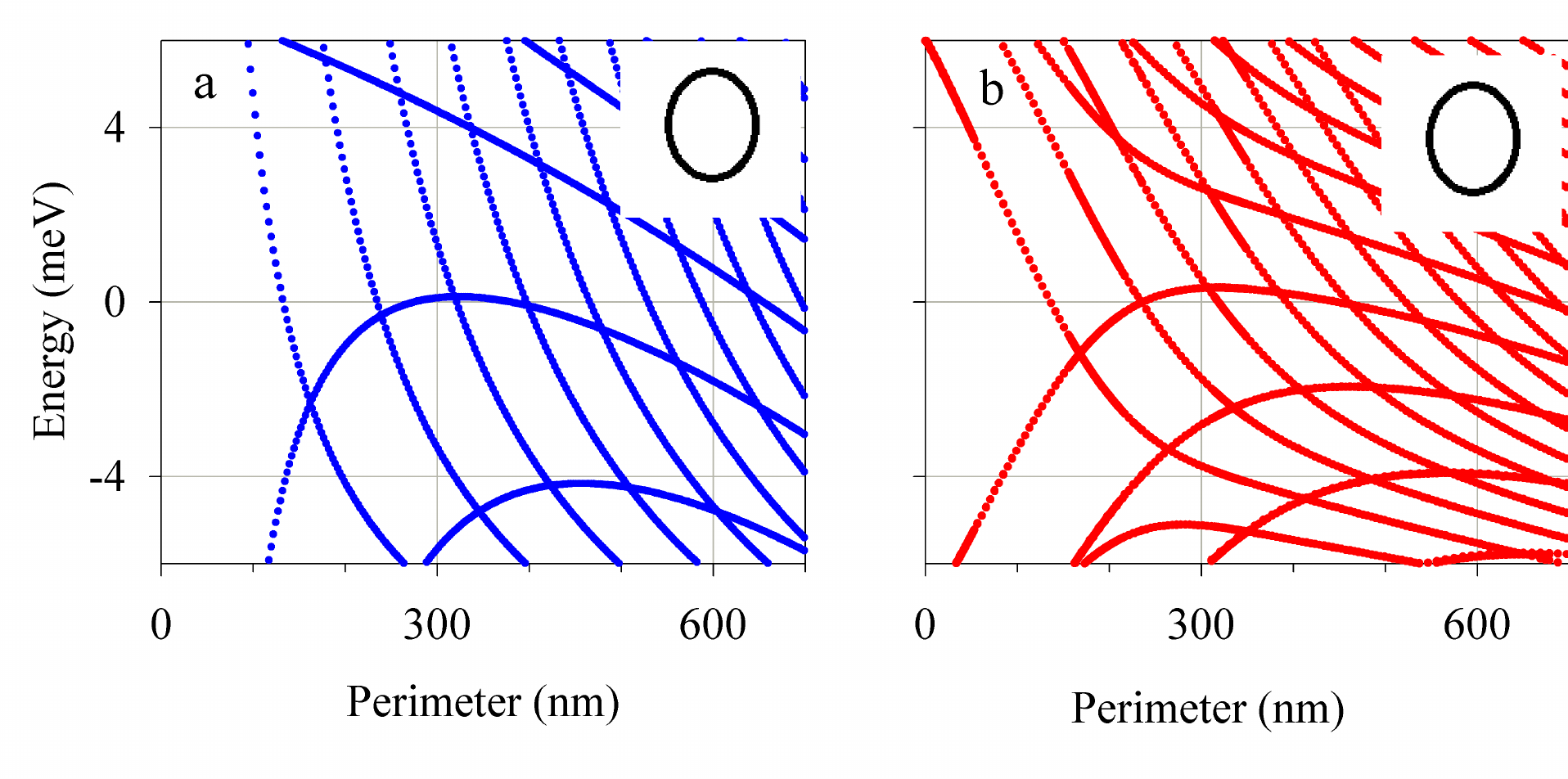}
\end{center}
 \caption{
Same as Fig.\ \ref{Fig2} with a finite magnetic field.
Parameters: wall potential $V_a=10\, {\rm meV}$, smoothness $s=12.5\, {\rm nm}$ and magnetic field $B=0.5\, {\rm T}$.
}
\label{Fig4}
\end{figure}

We discuss next the results in presence of a magnetic field $B$. Figure \ref{Fig4} compares the analytic and numeric results for a circular loop when $B=0.5$~T. The magnetic field 
breaks the valley degeneracy of the loop eigenmodes,
yielding asymmetric spectra with respect to energy inversion. 
We remark that Fig.\ \ref{Fig4} shows the results for a given valley.
The results for the opposite valley, not shown, are shifted
in the opposite direction restoring the symmetry of the spectrum for both energy and valley inversion (${\cal C}$ symmetry). There is again a good qualitative agreement between QPM and Q2DM, 
the most noticeable difference being the denser bunching of the energy branches within Q2DM, as in the $B=0$ cases discussed above.

\begin{figure}[t]
\begin{center}
\includegraphics[width=0.45\textwidth, trim = 0.5cm 10.5cm 1.cm 2.5cm,clip]{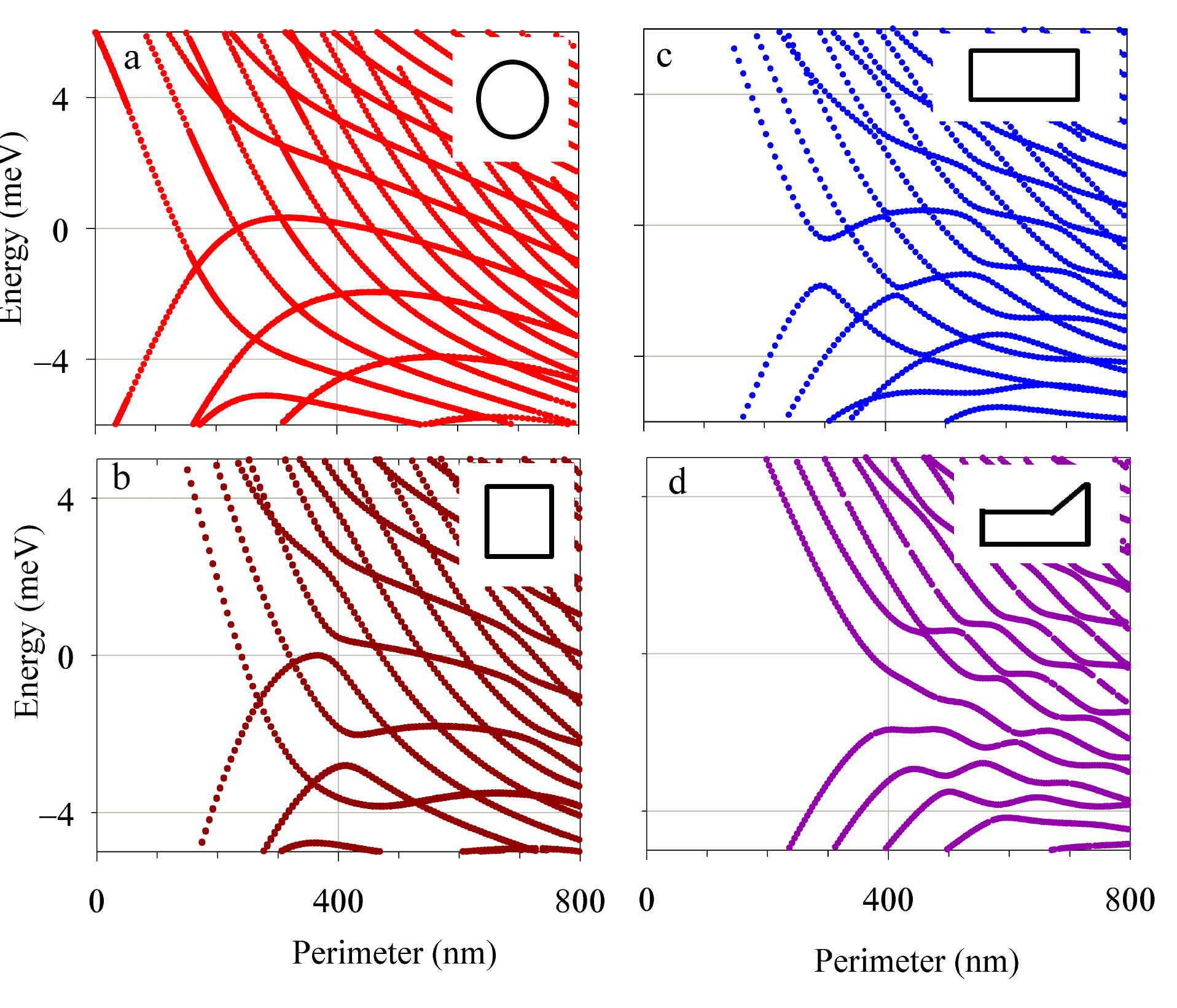}
\end{center}
 \caption{
Same as Fig.\ \ref{Fig3} with a finite magnetic field.
Parameters as in Fig.~\ref{Fig4}.
}
\label{Fig5}
\end{figure}

In Fig.\ \ref{Fig5} we show the results of the 2D model at finite magnetic field for the different shapes. 
The field-induded asymmetry is similar 
when the geometrical structure changes. Branch anticrossings are present for nonsymmetic shapes, although they are not centered at zero energy but shifted vertically. Similarly to $B=0$ (cf. Fig.\ \ref{Fig3}), we find that with an increase of the perimeter
the shape-dependent features of the spectrum are washed out. 

\begin{figure}[t]
\begin{center}
\includegraphics[width=0.45\textwidth, trim = 0.5cm 16.5cm 1.cm 2.5cm,clip]{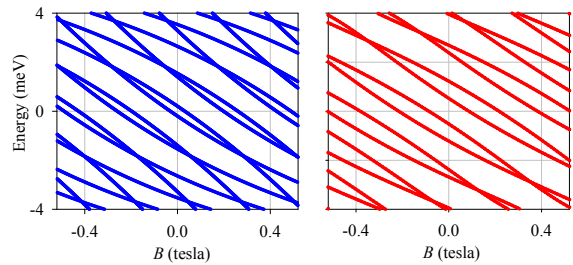}
\end{center}
 \caption{
Spectrum as a function of magnetic field for a square loop of a fixed perimeter in QPM (a) and Q2DM (b).
Parameters: wall potential $V_a=10\, {\rm meV}$, smoothness $s=12.5\, {\rm nm}$ and perimeter ${\cal P}=500\, {\rm nm}$.
}
\label{Fig6}
\end{figure}

As a function of magnetic field, for a fixed loop dimension and shape, the loop spectrum shows a periodic behavior that is reminiscent of the Aharonov-Bohm effect. Figure \ref{Fig6} shows the result for a square
in the QPM and Q2DM, both in good qualitative agreement. The energy branches come in doublets, originating from the two kink branches $f_{1,2}$ given by Eq.\ \eqref{eq1}. The results in Fig.\ \ref{Fig6} are in general asymmetric with respect to energy inversion, in the same way as for the results as a function of perimeter (Fig.\ \ref{Fig5}); the energy inversion symmetry being fulfilled at $B=0$ only. For different shapes the results show a similar periodicity, only that the field spacings between branch doublets must be scaled by the enclosed flux $\Phi$.

As with previous figures, the results of Fig.\ \ref{Fig6} are for a single valley; the energy branches for the reversed valley having the exactly opposite slope. The magnetic field slope of the branches is clearly indicating that the loop sustains persistent currents in finite fields, as discussed already for 
trivial and topological circular rings 
in Refs.\ \onlinecite{Recher07,xavier10}, respectively. Hence,
we investigate the shape dependence of the persistent current in topological loops as a possible probe of the geometric effects on topologically confined states. For definiteness, we focus on the current associated to the lowest positive-energy quasiparticle of the spectrum 
$E_1$ from the derivative
\begin{equation}
 J_1 = \frac{\partial E_1}{\partial B}\; .
\end{equation}

\begin{figure}[t]
\begin{center}
\includegraphics[width=0.4\textwidth, trim = 2.5cm 5.cm 1.cm 2.0cm,clip]{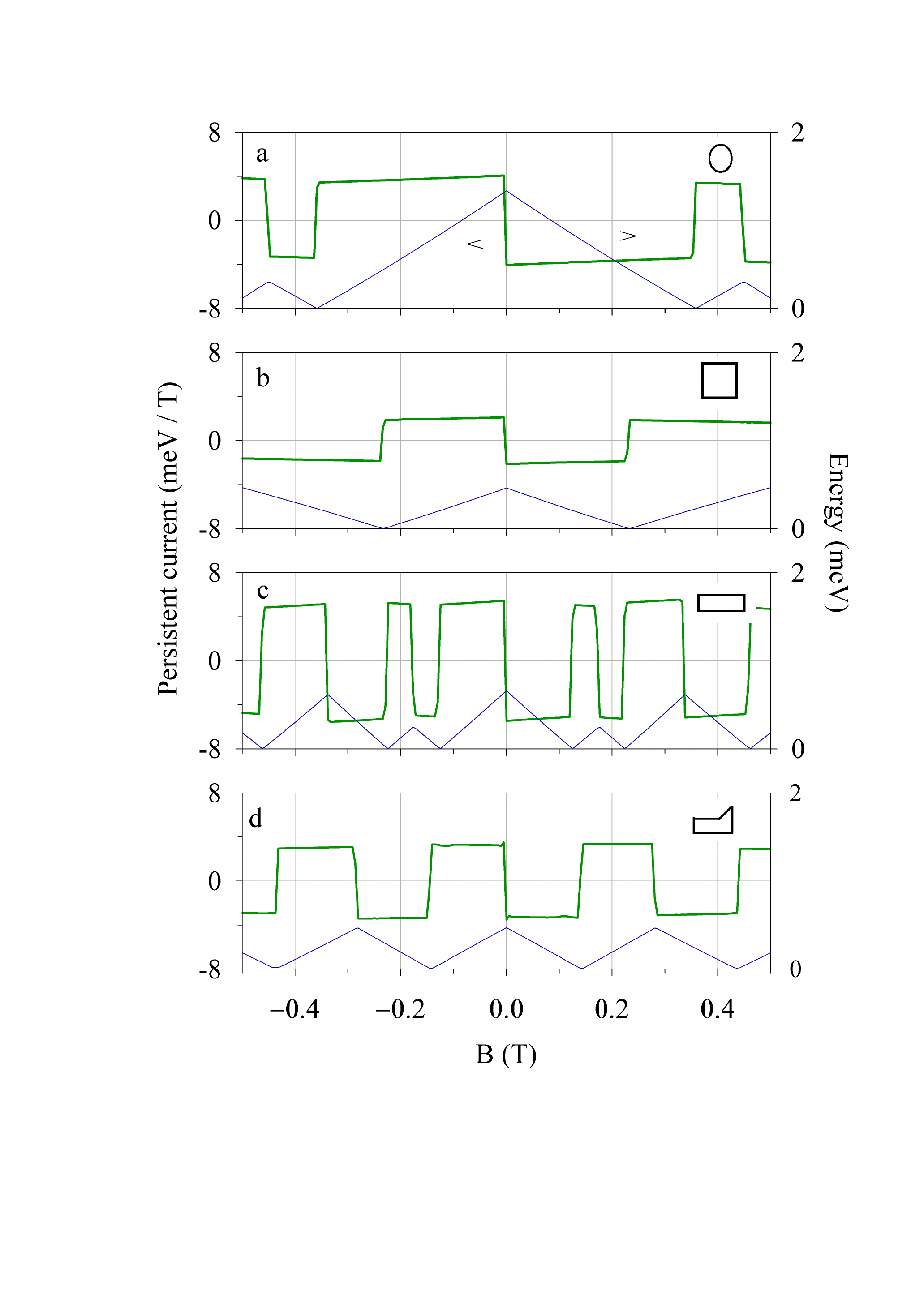}
\end{center}
\caption{
Persistent current $J_1$ (thick line, left scale) and energy $E_1$ (thin line, right scale)
of the lowest positive state as a function of the magnetic field for 
a fixed perimeter loop and the different shapes indicated in each panel.
Parameters as in Fig.~\ref{Fig6}.
}
\label{Fig7}
\end{figure}

The persistent current $J_1$ for the different shapes is shown in Fig.\ \ref{Fig7} for a fixed perimeter of 500 nm. Due to the aforementioned valley and energy inversion symmetry, the persistent current has an odd 
symmetry with respect to field inversion, implying a vanishing value at zero field. The current is characterized by a sequence of alternating plateaus of almost constant values. This is an effect of the nearly linear dispersion of the energy levels with magnetic field. Positive and negative plateaus
correspond to persistent currents of opposite valleys,
circulating with opposed chiralities. Along the sequence from high symmetry (a, circle) to low symmetry (c rectangle and d irregular shapes) the behavior looks qualitatively similar, but the plateau lengths are reduced due to the increased number of field transitions. This is also expected since the flux 
$\Phi$ is reduced for the deformed loops.
Therefore, measuring the persistent current yields valuable information on the geometrical structure of the loop.

\section{conclusions}

We have investigated the confined states in topological loops built in bilayer graphene.
We have devised an analytical model, the quantized perimeter model (QPM), based
on the infinite straight kink, that provides the wavelength quantization along the loop perimeter. In addition, a full quantum 2D model (Q2DM), valid at low energies, 
has been numerically solved in order to ascertain the validity of the analytical model. A general good agreement between both models is found. For large sizes, the energy spectra are almost insensitive to the loop shape, as expected from QPM. For small sizes, we have found that Q2DM reflects shape dependence in the emergence of zero-energy  crossings for circles, alternating crossings-anticrossings for squares, and only anticrossings for other 
more irregular structures. The magnetic field introduces energy-inversion asymmetries of the spectrum for a single valley and Aharonov-Bohm periodicities.
Shape-dependent anticrossings in small loops are also present with magnetic fields. Finally, we have found that persistent currents are sensitive
to shape-dependent features in presence of magnetic fields, which makes them a useful tool to look into bound states in topological systems.

Overall, our results are helpful for tailoring potential structures, finding the optimal geometry
for possible applications in quantum computation with topological qubits or for valleytronic devices. Since there could be
deviations from the intended symmetry due to fabrication limitations, our study emphasizes the importance of
taking into account emerging spatial asymmetries for a careful characterization of the underlying physical system.

\acknowledgments
We acknowledge support from AEI (Spain) Grant
No.\ PID2020-117347GB-I00, MINECO/AEI/FEDER Mar\'{\i}a de
Maeztu Program for Units of Excellence MDM2017-0711.

\appendix

\section{Modeling}

The Q2DM low energy Hamiltonian, Eq.\ (\ref{eq5}), requires a specific
spatial dependence of the asymmetric potential $V_a(x,y)$. This appendix contains 
the details of the modeling  of different loop shapes 
with $V_a(x,y)$ smooth profiles parameterized with a diffusivity $s$. The circle is described 
by a simple combination of logistic functions in the radial coordinate $r=\sqrt{x^2+y^2}$, 
\begin{eqnarray}
 V_a^{({\it cir})}(r) &=& V_a^{({\it in})}\, {\cal F}(r;R_0,s)\nonumber\\
&+&
 V_a^{({\it out})}\, \left(\,1-{\cal F}(r;R_0,s)\,\right)\; ,
\end{eqnarray}
where $V_a^{(in,out)}$ are the constant values of asymetric potential inside/outside of the loop and the logistic function reads
\begin{equation}
 {\cal F}(r;R_0,s)= \frac{1}{1+\exp{[(r-R_0)/s}]}\; .
\end{equation}

To describe smooth square and rectangle shapes we first introduce the ranges of the Cartessian coordinates $x\in[x_a,x_b]$, $y\in [y_a,y_b]$ 
representing the loop inner area
and then use a similar 
smooth parametrization in each Cartessian direction. Namely,
\begin{eqnarray}
\label{eqA3}
 V_a^{({\it squ,rect})}(r) &=& V_a^{({\it in})}\, {\cal F}_{2D}(x,y)\nonumber\\
 &+&
 V_a^{({\it out})}\, \left(\,1-{\cal F}_{2D}(x,y)\,\right)\; ,
\end{eqnarray}
where
\begin{eqnarray}
 {\cal F}_{2D}(x,y) &=& 
\left(\, {\cal F}(x;x_b,s)-{\cal F}(x;x_a,s) \,\right) \nonumber\\ 
&\times& \left(\, {\cal F}(y;y_b,s)-{\cal F}(y;y_a,s) \,\right)\; .
\end{eqnarray}
Finally, for the irregular shape we use the same modeling of Eq.\ (\ref{eqA3})
but with the modification that $y_b$ is no longer constant but the following
piecewise function of $x$
\begin{equation}
 y_b(x) = \left\{
 \begin{array}{cc}
0\;, &  x<0\; , \\
x\, y_b/x_b\;, &  x>0\; .
 \end{array}
 \right.
\end{equation}
Notice that we assume centered positions of the loop, with 
$x_{a,b}=\mp L_x/2$ and $y_{a,b}=\mp L_y/2$. 
For completeness, the perimeters for the different shapes are
${\cal P}_{\it cir}=2\pi R_0$,
${\cal P}_{\it squ}=4L_x$,
${\cal P}_{\it rect}=2(L_x+L_y)$,
${\cal P}_{\it irr}=\frac{3}{2}(L_x + L_y)+
\frac{1}{2}\sqrt{L_x^2+L_y^2}$.

\bibliography{loop}

\end{document}